\documentclass[11pt]{article}
	
\usepackage{amsmath}
\usepackage{amssymb}
\usepackage{graphicx}
\usepackage{setspace}
\usepackage{fullpage}
\usepackage{xcolor}
\usepackage{geometry}

\begin{document}

\title{Perturbations of lifespan inequality in natural populations}
\author{M.J. Wensink}
\date{University of Southern Denmark\\Department of Epidemiology, Biostatistics and Biodemography\\
Winsløwsvej 9B \\ 5000 Odense C \\ Denmark \\ mwensink@health.sdu.dk}
 
\maketitle

\section*{Keywords}
Inequality; lifespan; demography; sensitivity analysis; perturbation analysis

\section*{Competing interests}
Declarations of interests: none.

\section*{Abstract}
Correlations between high life expectancy and low lifespan inequality are frequently observed. A recent article seeks to explain this phenomenon by proposing that a mortality improvement maps to life expectancy and relative lifespan equality through the same weights $w(x)$, with an extra weight $W(x)$ applied for relative lifespan equality, specific for the equality indicator used. From statistical theory, the claim that changes of life expectancy and lifespan (in)equality map through the same weights is unexpected.

The current note explains this phenomenon by showing that $W(x)$ undoes (part of) $w(x)$. Thus, while the change in relative lifespan equality is proportional to the product of both weights, $w(x) \cdot W(x)$, it is proportional to neither $w(x)$ nor $W(x)$. As a result, some simplification is possible that gives way to an intuitive understanding and a more direct interpretation of the perturbation analysis proposed.

\newpage
\section{Introduction}
A pattern of correlated lifespan equality and life expectancy has been reported across extant primate species [1]. In human populations, historic increases in life expectancy tend to have coincided with increases in relative lifespan equality [2].  To explore why, Aburto et al. [2] explore the historical evolution of the \emph{sensitivities} of life expectancy and relative lifespan equality to mortality improvements: if a mortality reduction is achieved at age $x$, what is its effect on life expectancy and lifespan equality? The authors show that, historically, improvements have predominantly been made at ages at which a mortality reduction increased life expectancy and lifespan equality at the same time.

Aburto et al. [2] further seek to explain this historic link through mathematics: they propose that a mortality improvement maps to life expectancy and relative lifespan equality through the same weights $w(x)$, with an additional weight $W(x)$ applied for relative lifespan equality. This would seem to make the case that life expectancy and lifespan equality are mathematically linked (through $w(x)$). From statistical theory, this claim is unexpected.

The current note explains this phenomenon by showing that $W(x)$ undoes (part of) $w(x)$. Thus, while the change in relative lifespan equality is proportional to the product of both weights, $w(x) \cdot W(x)$, it is proportional to neither $w(x)$ nor $W(x)$. As a result, some simplification is possible that gives way to an intuitive understanding and a more direct interpretation of the perturbation analysis proposed in [2]. 

\section{The statistical point of view}
This section introduces notation and lays the groundwork for the later analysis. It contains standard probability theory that can be found in any textbook on the matter. See for example [3].

Statisticians concern themselves with random variables and their distributions. For a discrete random variable $X$, we wish to know the probability that $X$ takes the value $x$, i.e. $P(X=x)$. Those probabilities sum to 1. If a random variable is continuous (as in this note), $P(X=x)$ will always be 0, so instead we investigate the probability density function $f(x)$, which charts the intensity of occurrence of values sufficiently close to $x$. As for discrete random variables, $f(x)$ sums to 1 across the support of $X$ (domain of $f(x)$).

Furthermore,
\begin{equation}
P(X \leq x)= F(x),
\label{P(X<x)}
\end{equation}
where $F$ is the (cumulative) distribution function (cdf) of $X$. Because the cdf is defined for discrete and continuous random variables alike, i.e. it exists everywhere on $(\infty,\infty)$ and takes values in $[0,1]$, it is sometimes considered the fundamentally defining function of random variables [4]. For continuous random variables we then define
\begin{equation}
f(x):=\frac{dF(x)}{dx}.
\label{f(x)}
\end{equation}
Where confusion could arise, instead of $f(x)$ we may write $f_X(x)$, which reads as ``the pdf of $X$ parametrized in $x$''. Finally,
\begin{equation}
F(x)=\int_0^x f(a)da
\label{F(x)}
\end{equation}

If we wish to compute the mean of $X$ (its expected value), we solve:
\begin{equation}
E[X]=\int_{\mathcal{X}}x f(x) dx,
\label{E[X]}
\end{equation}
where $\mathcal{X}$ is the support of $X$.

We may also be interested in the variability of $X$ (about $E[X]$). A common measure is the variance:
\begin{equation}
VAR[X]=E[(X-E[X])^2]=E[X^2]-E[X]^2=\int_{\mathcal{X}}(x-E[X])^2 f(x) dx.
\label{VAR[X]}
\end{equation}
The variance, a measure of absolute inequality, is an \emph{expectation}: the $x$ in equation (\ref{E[X]}) is replaced by (the non-linear function) $(x-E[X])^2$.

More generally we might consider a transformation of $X$, say $Y=g(X)$. Since $X$ is a random variable, so is $Y$. In general, the mean of a transformation of a random variable does not equal the transformation of the mean. Rather, we have to do the transformation first and then calculate the mean:
\begin{equation}
E[g(X)]=\int_{\mathcal{X}}g(x) f(x) dx \neq g(E[X]).
\label{E[g(X)]}
\end{equation}
We will use this important fact to analyze the framework proposed by Aburto et al. [2].

With that, we have all the elements of standard statistics that are needed for our further analysis. We now continue to the demographer's view of the world, which tends to be formalized differently (often with good reason), but which does yield to standard statistics.

\section{The demographer's preoccupancy with rates}
Demographers tend to research in particular the \emph{timing} of events. Although some events may never happen, for example a couple may never have a third child, events like death are certain, and the interest lies on the \emph{when}, i.e. on the distribution of deaths across ages. In the statistical field of survival analysis, the random variable is therefore age (or time) at death (for humans usually measured in years), and we wish to know its distribution. Note that this random variable does not take negative values, since death is defined to occur after birth. Thus, just like in standard statistics we have $X$ with its pdf $f_X(x)$, which charts the distribution of deaths across ages.

Survivorship up to age $x$, commonly denoted $l(x)$, is that proportion of the population for which their age at death exceeds $x$:
\begin{equation}
l(x)=\int_x^{\infty}f(x)dx=1-\int_0^x f(x)dx=1-F_X(x).
\label{Survivorship}
\end{equation}
Differentiation of the far left and far right expression of equation (\ref{Survivorship}) yields
\begin{equation}
\frac{d l(x)}{dx}=-f(x)
\label{dl(x)}
\end{equation}
The density of deaths equals the rate of decline of $l(x)$.

Because a decline in $l(x)$ comes down only on those still alive at age $x$, demographers tend to focus on the mortality rate as the measure of intensity of death for those still alive:
\begin{equation}
\mu(x)=\frac{f(x)}{l(x)}=\frac{-l'(x)}{l(x)},
\label{Mortality Rate}
\end{equation}
the relative derivative of $l(x)$. The relevance of focusing on the mortality rate becomes clear if we consider centenarians. Suppose that we have few centenarians. This may be because centenarians are poor survivors, or because not many survive to become centenarians; we cannot tell. With mortality rates we can measure centenarians' survivability directly. Indeed, it has been suggested that one of the reasons we have so many female human centenarians but so few males is that generations of male centenarians have been wiped out due to smoking, leaving only very few male contestants [5]. This as such does not imply that male centenarians are poor survivors; a high mortality rate would.

Finally, the following relationships are of interest:
\begin{equation}
l(x)=\exp \{ -H(x) \} = \exp \left\{ -\int_0^x \mu(x) dx \right\}
\label{l(x)}
\end{equation}
and 
\begin{equation}
e(x)=\frac{\int_x^{\infty}af(a)da}{1-F(x)}=\frac{\int_x^{\infty} l(a)da}{l(x)},
\label{e(x)}
\end{equation}
Truncated mean $e(x)$ is generally referred to as remaining life expectancy at age $x$: given survivorship up to age $x$, what is the life expectancy from age $x$ onwards? When speaking of life expectancy, we usually refer to life expectancy at birth, $e(0)$, but we could calculate (remaining) life expectancy at any age. When mortality declines across ages, remaining life expectancy goes up; in a setting of constant mortality across ages, remaining life expectancy is invariant; in the case of increasing mortality, remaining life expectancy is a declining function of age. In populations with high mortality at the youngest and highest ages, such as in most human populations, remaining life expectancy goes up in the first years of life, but then goes down as the aging process takes hold.

The functions $\mu(x)$, $H(x)$, $l(x)$, $f(x)$, and $F(x)$ each give complete information about the distribution of deaths: knowing any one of these allows computation of all the others through the relationships given above. Demographers, while fully aware of standard statistics, tend to focus on $H(x)$, $l(x)$, $\mu(x)$, but results for $f(x)$ or $F(x)$ carry over to $\mu(x)$, $H(x)$, and $l(x)$ through their standard mathematical relationships without reservation.

\section{Inequality measures as expectations}
Aburto et al. [2] use three measures of relative equality all of which are derived from absolute \emph{inequality} (variability) measures. Absolute inequality is first divided by life expectancy to make it relative. Then the natural logarithm is taken to assure that the values that the indicators might take are unconstrained. Finally, Aburto et al. [2] take the additive inverse to assure that the resulting number is a measure of \emph{equality} rather than inequality. Let $A$ be a measure of absolute lifespan inequality and $R$ be a measure of relative equality, then
\begin{equation}
R=-\ln\left( \frac{A}{e_0} \right)
\label{AbsRel}
\end{equation}

The choice of $A$ depends on how we weigh differences. The variance (equation (\ref{VAR[X]}), Aburto et al. [2] use its square root, the standard deviation) is the mean squared difference from the mean. This means that large deviations from the mean are counted heavier than small deviations from the mean. To account in such a way is a modeling choice. We might also consider the mean absolute difference between two points, where larger differences count the same as smaller differences. The mean absolute distance within pairs drawn from the same distribution is the numerator of the Gini coefficient ([2], see also below):
\begin{equation}
G=\frac{E[|X_i-X_j|]}{E[X]},
\label{G}
\end{equation}
where $X_i$ and $X_j$ have the same distribution.

Alternatively, we might consider the expected number of life years lost at death:
\begin{equation}
e^{\dagger}=E[e(x)]=\int_0^{\infty}e(x)f(x)dx.
\label{E dagger}
\end{equation}
This latter measure is known as the life table entropy [7-10]. The rationale for using this measure is that if all die at the same time, no life years are lost at death, since putting the dead back into the population would lead to renewed death at the exact same time, and $e^{\dagger}$ equals zero. Conversely, if putting the dead back into the population tends to lead to renewed death at a very different age, this means that deaths are very much dispersed across ages, and $e^{\dagger}$ is large.

While Aburto et al. [2] parametrize these measures in terms like $\mu(x)$, $H(x)$, and $l(x)$, all of these measures can be written as expectations without loss of generality: $E[(X-E[X])^2]$ (or its square root), $E[|X_i-X_j|]$, and $E[e(x)]$.

\section{Perturbations in survival analysis}
We may wish to explore the way life expectancy or measures of (in)equality change in response to infinitesimal perturbations in some function at age $x$ (technically: around age $x$, taking the limit as the perturbation area goes to 0, see [11]), using calculus to express these sensitivities as derivatives or to write the measures in a differential form. We first have to decide which function we are going to perturb, and how. We could perturb mortality $\mu(x)$: if mortality changed at age $x$, what would the result be on the measure? But we may also choose to perturb any of the functions $f(x)$, $H(x)$, $l(x)$ or $F(x)$. It is reasonable to model mortality perturbations because these can be interpreted as a change in survivability at age $x$, uncoupled from events at other ages (see above). Furthermore, perturbations could be additive (as in standard calculus), but they might also be proportional. This is essentially a matter of parametrization: any modeling choice goes, as long as one is consistent (see [11]). Aburto et al. choose parametrization $\rho(x)=(-d\mu(x)/dx)/mu(x)$; a reasonable choice. It means that $-\rho(x)\mu(x)dx=d\mu(x)$, which takes us back to standard calculus.

Although survival analysis is simply the analysis of the distribution of survival times, the directionality of time does change a thing or two. Given a distribution of body heights, if we were told that, through some miraculous intervention, somebody no longer has body height $x$, the new body height could be higher or lower. However, if somebody's life is saved at age $x$, the directionality of time tells us that the survival time of that person will now exceed $x$. This makes survival analysis subject to a perturbation analysis that does not usually apply in other fields of statistics. As a result, such perturbation analysis tends to be framed in terms of demographic functions like $\mu(x)$ or $l(x)$, but we can switch between standard statistics and demography (survival analysis) flexibly and without loss of generality, realizing that lives saved at age $x'$ are redistributed according to the left-truncated distribution of $X$, that is, $f(x)$ to the right of $x'$.

\subsection{Perturbation of life expectancy}
Now what would the effect be of a mortality perturbation at age $x$ on life expectancy [10,12,13]? If we have a perturbation parameter $\varepsilon$, a generic placeholder for a perturbation,
\begin{equation}
\frac{d e_0}{d \varepsilon}=\int_0^{\infty} \frac{d l(x)}{d \varepsilon}
\label{d e_0}
\end{equation}
For an isolated mortality perturbation at age $x$, this reduces to
\begin{equation}
\frac{d e_0}{d \varepsilon}=\int_{x}^{\infty} \frac{d l(a)}{d \varepsilon}da
\label{Perturb e_0}
\end{equation}
In words, the total change equals the change of the respective integral to the right of $x$. Wrycza and Baudisch [13] express this as 
\begin{equation}
-\frac{d\mu(x)}{d\varepsilon}l(x)e(x);
\label{Wrycza e_0}
\end{equation}
Aburto et al. [2] choose to express this as 
\begin{equation}
\rho(x)\mu(x)l(x)e(x)=\rho(x)w(x).
\label{Aburto e_0}
\end{equation}
Notice that $l(x)e(x)$ is the area under the curve to the right of $x$. Thus, it makes sense to see $e(x)$ (or indeed $l(x)e(x)$) here: $e(x)$ is the mean (expected) shift in the age of death of those lives saved at $x$:
\begin{equation}
e(x')=E_{x'}[X],
\label{eq:}
\end{equation}
(the symbol $x'$ is used to explicitly refer to truncated distributions (truncated at age $x'$)). However, since the measures of absolute inequality considered in [2] are non-linear functions of $X$, and since the expectation of a transformation of a random variable is not the transformation of the expectation (inequality (\ref{E[g(X)]})), it is unclear why perturbations should be mapped through $l(x)e(x)$ to find their effect on measures of (in)equality. While $l(x)$ could take the role of a normalizing constant, the occurrence of $e(x)$ is particularly problematic because it seems to violate the rule that to find a mean of a transformation, we need to do the transformation first and then calculate its mean, rather than the other way round.

\section{Relative measures of inequality and their perturbation}
If we perturb absolute inequality $A$ relative to perturbation parameter $\varepsilon$, realizing that $d\ln(x)/dx=1/x$ and applying the chain rule, we get
\begin{equation}
\frac{dR}{d \varepsilon}=\frac{d e_0/d \varepsilon}{e_0}-\frac{d A / d\varepsilon}{A}.
\label{dR}
\end{equation}
In [2] we find this form in equations (3) and (10). We do not immediately see it in equation (7) because the Gini coefficient is here expressed in an alternative form (see below). Expression (\ref{dR}) means that we can separately analyze the effects of a mortality perturbation on $A$ versus its effect on $e_0$; the result is a linear combination of the relative derivative of each.

Because measures of relative equality all are measures of absolute inequality divided by life expectancy and then processed further, it may seem convenient to have weights $w(x)$ in expression (\ref{dR}): this will get the perturbation of the life expectancy component right. However, since the effect of a perturbation on the $A$ part of $R$ is not expected to be weighted by $e(x)$ because in general $E[g(X)] \neq g(E[X])$, the weight $W(x)$ may have to ``unweigh'' by $l(x)e(x)$ for the $A$-part of the expression. So what is the role of $l(x)e(x)$ in each of the measures considered in [2]? 

\subsection{The coefficient of variation (and a variant)}
Would a change in variance (or its square root) in response to a mortality perturbation at age $x'$ be proportional to $l(x')e(x')$? It is unclear why. The deaths happening at age $x'$ are redistributed to $x>x'$ with expectation $e(x')$, which has a linear effect on $E[X]$. The final effect on $E[(X-E[X])^2]$, however, depends on the interaction of the change in $E[X]$ with the particular shape of the distribution truncated at $x'$.

Looking at the mathematics, we find the following. Aburto et al. [2] find the weight $W_v(x)$ for (the negative log of) the coefficient of variation to be
\begin{equation}
W_v(x)=\frac{1}{e_0}-\frac{CV(x)}{\sigma^2}.
\label{W_v(x)}
\end{equation}
Notice that $d\sigma / d \varepsilon = (d \sigma^2 /d\varepsilon )/ 2\sigma$, so $d\sigma/d\varepsilon=CV(x)/\sigma$ hence $\sigma^2$ in equation (\ref{W_v(x)}). Noting the equations that follow equation (12) of Aburto et al. [2], further developed as equation (C1) in their Appendix, we find that 
\begin{equation}
CV(x)=\frac{\int_x^{\infty}l(a)(a-e_0)da}{l(x)e(x)}.
\label{CV(x)}
\end{equation}
Hence, $W_v(x)$ unweighs by $l(x)e(x)$, for the entire part of the expression that pertains to $A$. What remains is $\int_x^{\infty}l(a)(a-e_0)da$, which involves the exact shape of the truncated distribution, not just its expected value.

\subsection{The Gini coefficient (and a variant)}
The Gini coefficient [6] requires more introduction. It is a measure of inequality commonly used in economics to measure income inequality, but there is no reason why it could not be used to measure inequality in anything else. In particular, lifespans are necessarily positive (as are gross incomes), making survival analysis a natural area of application of the Gini index [14]. The standard form of the Gini coefficient is 
\begin{equation}
G=\frac{1}{2 E[X]}\int_0^{\infty}\int_0^{\infty}|x_1-x_2|f(x_1)f(x_2)dx_1dx_2,
\label{Standard Gini}
\end{equation}
where $x_1$ and $x_2$ are dummy variables for $x$ to keep track of the order of integration. The division by $2E[X]$ arises because the rest of the expression counts the differences within every pair of values of $X$ twice: once through $x_1$, once through $x_2$. It is common to define the Gini coefficient as the mean of the absolute differences ($\int_0^{\infty}\int_0^{\infty}|x_1-x_2|dF(x_1)dF(x_2)$) divided by twice the mean, although perhaps it is more just to say that the double integral $\int_0^{\infty}\int_0^{\infty}|x_1-x_2|dF(x_1)dF(x_2)$ is the mean of twice the absolute differences (as it counts every pair twice). Thus, as the numerator of the Gini coefficient we have
\begin{equation}
E[|X_1-X_2|],
\label{Numerator Gini}
\end{equation}
where $X_1$ and $X_2$ have the same distribution. The Gini coefficient as a whole is then a relative measure of inequality (because divided by the expectation of $X$, which in our setting is life expectancy $e_0$), and Aburto et al. [2] continue by taking the additive inverse of its natural logarithm.

The Gini coefficient can be rewritten in many ways (Ceriani and Verme [15] find no fewer than 13 equivalent expression in Gini's original work alone). Hanada [16] is the usual reference for the re-expression below in demography, but Lubrano [17] contains some very instructive reworkings. First, Lubrano states that realizing that $F(x)$ and $1-F(x)$ are the proportions of individuals with [lifespans] below and above $x$, we can simply integrate the product of these probabilities across the domain of $X$, which gives the Gini coefficient in its form
\begin{equation}
G=\frac{1}{E[X]}\int_{\mathcal{X}}F(x)(1-F(x))dx.
\label{Alternative Gini}
\end{equation}
Then, realizing that $|x_1-x_2|=(x_1+x_2)-2\min(x,y)$ and that the distribution of the $\min$ of two random variables that have the same distribution is $1-(1-F(x))^2$, after some reworking (details in [17]), we obtain the expression used by Aburto et al. [2], 
\begin{equation}
G=1-\frac{\vartheta}{e_0},
\label{Gini Aburto}
\end{equation}
where $\vartheta=\int_0^{\infty}l(x)^2dx$.

Would the change in $E[|X_1 - X_2|]$ in response to a mortality perturbation at age $x'$ be proportional to $l(x')e(x')$? The absolute difference between any pairs $(x_1,x_2)$ such that $x_1 < x'$ and $x_2 < x'$ would not change. Similarly, the absolute difference between any pairs $(x_1,x_2)$ such that $x_1 > x'$ and $x_2 > x'$ existing before the perturbation would not change. What \emph{would} change is the difference between any pairs $(x_1,x_2)$ where $x_1=x'$ or $x_2=x'$ or both. If a pair consists of $x'$ and some $x<x'$, we expect the difference on average to increase by $l(x')e(x')$. Hence, part of the change in $E[|x_1 - x_2|]$ \emph{will} be proportional to $l(x')e(x')$. However, for pairs consisting of $x'$ and some $x \geq x'$, anything can happen: $x'$ may be redistributed to a point smaller than, equal to, or greater than $x$, and there is no simple relationship that makes us expect that the result be proportional to $l(x')e(x')$.

It is not straightforward to determine intuitively what would happen to the numerator of the Gini coefficient, so we resort to mathematics. Rewriting $G$,
\begin{equation}
G=1-\frac{\vartheta}{e_0}=\frac{e_0}{e_0}-\frac{\vartheta}{e_0}=\frac{e_0-\vartheta}{e_0},
\label{Gini Wensink}
\end{equation}
we see that $e_0-\vartheta = E[|X_1-X_2|]$. Differentiating we find that
\begin{equation}
\frac{d (e_0-\vartheta)}{d\varepsilon}=\frac{d e_0}{d\varepsilon}-\frac{d \vartheta}{d\varepsilon}=e(x)-\frac{d \vartheta}{d\varepsilon}.
\label{d E[|x_1-x_2|]}
\end{equation}
Aburto et al. find the following expression for $d\vartheta/d\varepsilon$ [2]:
\begin{equation}
\frac{d\vartheta}{d\varepsilon}=\rho(x)\mu(x)l(x)e(x)2 l(x)\bar{l}(x),
\label{d tuvle Aburto}
\end{equation}
where
\begin{equation}
\bar{l}(x)=\frac{1}{l(x)}\int_x^{\infty} c_x(a)l(a)da
\label{l bar}
\end{equation}
From equation (B1), the unnumbered equation just below (B1), and the fact that
\begin{equation}
c(x)_{x'}=l(x)/\int_{x'}^{\infty}l(a)da=l(x)/(l(x')e(x')),
\label{c(x)_x}
\end{equation}
i.e., the life table age composition conditioned on truncation at age $x'$, a clarified version on the equation that can be found on page 8 of [2], we find that
\begin{equation}
\frac{d \vartheta}{d\varepsilon} = \frac{d\mu(x)}{d\varepsilon}\int_x^{\infty} l(a)^2da,
\label{d tuvle me}
\end{equation}
which does not involve $l(x)e(x)$. Again, $l(x)e(x)$ in $w(x)$ was canceled out.

\subsection{The life table entropy (and a variant)}
The life table entropy $\bar{H}$ is a measure of relative variation in the length of life, defined as 
\begin{equation}
\bar{H}=\frac{\int_0^{\infty}l(x)\ln(l(x))dx}{\int_0^{\infty}l(x)dx}=\frac{e^{\dagger}}{e_0}
\label{H bar}
\end{equation}
In this equation, $e^{\dagger}$ is a measure of absolute inequality in lifespans. We can rewrite it as
\begin{equation}
e^{\dagger}=E[e(x)]=\int_0^{\infty}f(x)e(x)dx
\label{e dagger}
\end{equation}
(see [9,10]). Thus, $e^{\dagger}$ signifies the wonderfully paradoxical quantity ``expected life expectancy at death''. The relationship with the effect of mortality perturbations on life expectancy may seem direct: by reducing mortality at age $x'$, deaths are redistributed to ages $x>x'$ with expectation $e(x')$. But would we expect it to be proportional to $e(x')$?

If deaths are redistributed from age $x'$ to ages $x>x'$ with expectation $e(x')$, all deaths that occurred before $x'$ see their remaining life expectancy at death increase proportional to $e(x')$. 

Furthermore, the effect of the perturbation will be a negative effect proportional to $e(x')$ for the deaths that no longer take place at $x'$.

Finally, because more deaths will now take place above age $x'$, the contributions of $e(x)$ for $x > x'$ have to be re-weighted. We therefore expect to see a positive effect proportional to 
\begin{equation}
\int_{x'}^{\infty} e(x) f(x) dx / l(x')=e^{\dagger}(x'),
\label{E dagger x}
\end{equation}
which is not related to $e(x')$, at least not through a straightforward relationship.

Indeed, this is precisely what we see. The weight for $h$, a relative measure of equality based on $e^{\dagger}$, is
\begin{equation}
W_h(x)=\frac{1}{e_0}-\frac{H(x)+\bar{H}(x)-1}{e^{\dagger}}
\label{W_h(x)}
\end{equation}
Notice that $H(x)=\int_0^x \mu(a)da$ sums over ages up to $x$, while
\begin{equation}
\bar{H}(x)=\frac{\int_x^{\infty}f(a)e(a)da}{\int_x^{\infty}l(a)da}=\frac{e^{\dagger}(x)}{l(x)e(x)}
\label{H bar x}
\end{equation}
pertains to ages from $x$ onwards. From equation (\ref{H bar x}) we see that the weight $W_h(x)$ cancels out $l(x)e(x)$ for exactly that part of $e^{\dagger}$ for which the effect of a mortality perturbation is not proportional to $l(x)e(x)$. In sum, the effect of a mortality perturbation on $\bar{H}$ is a mix of effects that are proportional to $l(x)e(x)$ with effects that are not.

\section{Discussion}
Given that Aburto et al. [2] aimed to analyze changes in lifespan and relative lifespan equality in a single framework, including $w(x)$ in the sensitivities of indicators of equality is a convenient modeling choice; this will automatically get the change in the life expectancy part right. However, $w(x)$ decomposes into a part that is there by construction, $\mu(x)$, which takes us back from $\rho(x)$ to standard calculus, versus a part $l(x)e(x)$ that had to be canceled out against $1/(l(x)e(x)$ in (part of) $W(x)$. This seems to defeat the purpose of weighing. In generalized linear models, contributions of groups of observations to the likelihood are weighed by group size, which has a direct interpretation. In weighted means of a population of sub populations, we would weigh by the respective sub population sizes to account for the importance of their contribution to the grant total. In Mantel-Haenzel weights of stratified analyses, the weights reflect the amount of information in each stratum. What is the interpretation of $w(x)$ and $W(x)$? Weights $w(x)$ have an interpretation only in the context of life expectancy changes, but not for changes in lifespan equality, while $W(x)$ has no interpretation on its own. Rather, we could parametrize in $d\mu(x)$, with single, different weights for life expectancy and the various indicators of lifespan equality. Thus, the interpretation is problematic. The effect of a change in mortality at age $x$ is proportional to $w(x)W(x)$, but not to $w(x)$ and not to $W(X)$. By clarifying the expressions derived in [2], it is possible to arrive at a more intuitive understanding of the way measures of inequality respond to mortality perturbations. For each of the measures we were able to understand, in words, what the effect of a perturbation would be, and why it would not generally be proportional to $l(x)e(x)$, i.e. not to $w(x)$.

A further potential drawback of studying life expectancy and relative equality through the same framework is that it obscures whether any increase in relative equality comes from increasing life expectancy (keeping absolute inequality constant), or from declining absolute inequality (keeping life expectancy constant). Again, whether or not we find this a disadvantage depends on whether or not we consider the former case progress, and our modeling choice follows from that. In practice it may often be a little bit of both [18]. But economists and social scientists make a point of separating averages (means) from the spread around that mean in their research. There is an increasing focus on the fact that if a society as a whole is well off, large subgroups may nevertheless not exceed subsistence levels (see e.g. [19]). In human demography, the same is true for inequalities in life expectancy: if societies as a whole experience greater life expectancy, this does not imply that all subgroups are better off [20-23], even leading to the study of ``deaths of despair'' [24]. Hence, even though increasing life expectancy and decreasing lifespan inequality are often co-joined, especially in populations where mortality increases with age, it is vital to study them separately: if they are co-joined, there is little to study, while it often draws our interest, and indicates important social issues, when they are not. The investigation of two measures that are not correlated by construction, i.e. life expectancy and some absolute measure of inequality, thus gives a clearer view.

Sensitivity measures can often be split up in the product of something that measures presence, i.e., some measure of the probability that an individual is affected by some perturbation, multiplied by the impact of the perturbation on the affected individuals. Indeed, it seems that this is generally the case when the system we are studying is rooted in probability theory. For example, Caswell [25,26] showed that in (evolutionary applications of) stable population theory, the sensitivity of the population growth rate $r$ to age-specific changes in mortality can be written as the product of the population age distribution (presence) and reproductive value (impact). The reproductive value charts the expected reproductive contribution of an organism from age $x$ onwards given that it has survived up to age $x$ (see also [27]).

In the present setting, no reproduction is involved and no population growth rate is investigated; the system is in many ways simpler than stable population theory. However, it may be harder to decide what our measure of presence would be. Would it be survival $l(x)$ (``those who are there''), would it be the life table age distribution $c(x)$ (``those who are relatively there''), or would it be the distribution of deaths, $f(x)$? All of these are reasonable choices ($c(x)$ would be suggested by stable population theory), and changing between one and the other is done by simply moving one term from ``impact'' to ``presence'' or vica versa. A recasting in such terms may aid interpretation.

A future direction of research worth mentioning is the modeling of unobserved heterogeneity [28]. The idea here is that, just like for variables that we do observe, there are unobserved factors that we can collapse into one variable that works multiplicatively on mortality. Suppose that we have a distribution for cholesterol that does not change over lifetime. Those with high (or very low, but we will ignore that for the moment) cholesterol tend to die first. Similarly, if we have unmeasured heterogeneity that we think of as one variable, which we shall call frailty, those who die are expected to have a higher frailty value than those who survive. In the same vein, any people that we save will tend to have a higher measure of frailty than those who did not need to be saved. But how much higher? Mathematical relationships exist [28], and although this goes beyond the scope of the current paper, it would be interesting to expand the perturbation analysis lifespan inequality for this insight. Since we are saving frailer people, their remaining life expectancy will be lower, and their impact on measures of inequality will differ as well.

\section{Conclusion}
While the approach of weighing by $w(x)$ and $W(x)$ developed in [2] seems unnecessary, the research initiated in [2] gives scope for ample new research. This note explored how the weights proposed in [2]arise and simplify, and, if successful, provides an intuitive understanding of the way absolute measures of inequality change as a result of mortality perturbations. While measures of inequality tend to be correlated [29], we may find situations where these measures \emph{disagree}, in particular under selection of unobserved heterogeneity, which are suggested as avenues of future research.

\newpage
\small
\section{References}

[1] F. Colchero, R. Rau, O. R. Jones, J. A. Barthold, D. A. Conde, A. Lenart, L. Nemeth, A. Scheuerlein, J. Schoeley, C. Torres, V. Zarulli, J. Altmann, D. K. Brockman, A. M. Bronikowski, L. M. Fedigan, A. E. Pusey, T. S. Stoinski, K. B. Strier, A. Baudisch, S. C. Alberts, and J. W. Vaupel, The emergence of longevous population," Proc. Nat. Aca. Sci., vol. 113, pp. E7681-E7690, 2016.\\

[2] J. M. Aburto, F. Villavicencio, U. Basellini, S. Kjærgaard, and J. W. Vaupel, Dynamics of life expectancy and life span equality," Proc. Nat. Aca. Sci., vol. 117, pp. 5250-5259, 2020.\\

[3] I. Miller and M. Miller, John E. Freund's Mathematical Statistics with Applications, 8th edn. Pearson, 2014.\\

[4] G. Grimmett and D. Stirzaker, Probability and Random Processes, 3rd edn. Oxford University Press, 2001.\\

[5] T. T. Perls, Male centenarians: How and why are they different from their female counter-parts?, J Am Geriatr Soc., vol. 65, pp. 1904-1906, 2017.\\

[6] C. Gini, Variabilità e mutabilità, in Memorie di Metodologica Statistica, E. Pizetti, T. Salvemini, Eds. Libreria Eredi Virgilio Veschi, Rome, 1912.\\

[7] C. E. V. Leser, Variations in mortality and life expectation, Population Studies, vol. 9,pp. 67-71, 1955.\\

[8] N. Keyfitz, Introduction to the Mathematics of Population: With Revisions. Addison-Wesley Series in Behavioral Science: Quantitative Methods, Addison-Wesley, 1977.\\

[9] N. Goldman and G. Lord, A new look at entropy and the life table., Demography, vol. 23, pp. 275-282, 1986.\\

[10] J. W. Vaupel, How change in age-specific mortality affects life expectancy, Population Studies (Camb.), vol. 40, pp. 147-157, 1986.\\

[11] M. J. Wensink, T. F. Wrycza, and A. Baudisch, No senescence despite declining selection pressure: Hamilton's result in broader perspective, Journal of Theoretical Biology, vol. 347, pp. 176-181, 2014.\\

[12] H. Caswell, Perturbation analysis of nonlinear matrix population models, Demographic Ressearch, vol. 18, pp. 59-116, 2008.\\

[13] T. F. Wrycza and A. Baudisch, How life expectancy varies with perturbations in age-specific mortality, Demographic Research, vol. 27, pp. 365-376, 2012.\\

[14] M. Bonetti, C. Gigliarano, and P. Muliere, The gini concentration test for survival data, Lifetime Data Anal., vol. 15, pp. 493-518, 2009.\\

[15] L. Ceriani and P. Verme, The origins of the gini index: extracts from variabilità e mutabilità (1912) by corrado gini, J Econ Inequal, vol. 10, pp. 421-443, 2012.\\

[16] K. Hanada, A formula of gini's concentration ratio and its application to life tables, Journal of the Japanese Statistical Society, vol. 13, pp. 95-98, 1983.\\

[17] M. Lubrano, The econometrics of inequality and poverty. Chapter 4: Lorenz curves, the Gini coefficient and parametric distributions. Lecture notes, Aix-Marseille Université, site de la Vieille Charité à Marseille, 2017.\\

[18] J. M. Aburto, M. Wensink, A. van Raalte, and R. Lindahl-Jacobsen, Potential gains in life expectancy by reducing inequality of lifespans in denmark: An international comparison and cause-of-death analysis., BMC Public Health, vol. 18, p. 831, 2018.\\

[19] T. Piketty, Capital in the Twenty First Century, English edn. Harvard University Press, 2014.\\

[20] J. M. Aburto and A. van Raalte, Lifespan dispersion in times of life expectancy fluctuation: The case of central and eastern europe., Demography, vol. 55, pp. 2071-2096, 2018.\\

[21] A. A. van Raalte, I. Sasson, and P. Martikainen, The case for monitoring life-span inequality, Science, vol. 362, pp. 1002-1004, 2018.\\

[22] I. Sasson, Trends in life expectancy and lifespan variation by educational attainment: United States, 1990-2010, Demography, vol. 53, pp. 269-293, 2016.\\

[23] J. W. Vaupel, Z. Zhang, and A. A. van Raalte, Life expectancy and disparity: An international comparison of life table data, BMJ Open, vol. 1, p. e000128, 2011.\\

[24] A. Case and A. Deaton, Deaths of Despair and the Future of Capitalism. Princeton University Press, 2020.\\

[25] H. Caswell, A general formula for the sensitivity of population growth rate to changes in life history parameters, Theoretical Population Biology, vol. 14, pp. 215-230, 1978.\\

[26] H. Caswell, Reproductive value, the stable stage distribution, and the sensitivity of the population growth rate to changes in vital rates, Demographic Research, vol. 23, pp. 531-548, 2010.\\

[27] M. J. Wensink, H. Caswell, and A. Baudisch, The rarity of survival to old age does not drive the evolution of senescence," Evol. Biol., vol. 44, pp. 5-10, 2017.\\

[28] J. W. Vaupel and T. Missov, Unobserved population heterogeneity: A review of formal relationships, Demographic Research, vol. 31, pp. 659-688, 2014.\\

[29] A. A. van Raalte and H. Caswell, Perturbation analysis of indices of lifespan variability, Demography, vol. 50, pp. 1615-1640, 2013.

\end{document}